\documentstyle[epsfig]{mn}

\title{Population III Star Clusters in the Reionized Universe}

\author[J.L. Johnson]
{Jarrett L. Johnson\thanks{E-mail: jjohnson@mpe.mpg.de} \\
Max-Planck-Institut f{\"u}r extraterrestrische Physik, 
Giessenbachstra\ss{}e, 85748 Garching, Germany \\
Theoretical Modeling of Cosmic Structures Group \\}

\pubyear{2009}

\begin{document}
\maketitle
\topmargin-1cm

\begin{abstract}
In reionized regions of the Universe, gas can only collapse to form stars in dark matter (DM) haloes which grow to be sufficiently massive.  
If star formation is prevented in the minihalo progenitors of such DM haloes at redshifts $z$~$\ga$~20, then these haloes 
will not be self-enriched with metals and so may host Population (Pop) III star formation.  We 
estimate an upper limit for the abundance of Pop III star clusters which thus form in the reionized Universe,
as a function of redshift.  Depending on the minimum DM halo mass for star formation, between of the order of one and of the order of a thousand 
Pop~III star clusters per square degree may be observable at 2 $\la$ $z$ $\la$ 7.  Thus, there may be a sufficient 
number density of Pop~III star clusters for detection in surveys such as the Deep-Wide Survey (DWS) to be conducted by the {\it James Webb Space Telescope} (JWST). 
We predict that Pop~III clusters formed after reionization are most likely to be found at $z$ $\ga$ 3 and within $\sim$ 40 arcsec ($\sim$ 1 Mpc comoving) 
of DM haloes with masses of $\sim$ 10$^{11}$ M$_{\odot}$, the descendants of the haloes at $z$ $\sim$ 20 which host the first galaxies that begin reionization.
However, if star formation is inefficient in the haloes hosting Pop~III clusters due to the photoionizing background radiation, these clusters may not
be bright enough for detection by the Near-Infrared Camera which will conduct the DWS.  Nonetheless, if the stellar initial mass function (IMF) is top-heavy 
the clusters may have sufficiently 
high luminosities in both Ly$\alpha$ and He~{\sc ii}~$\lambda$1640 to be detected and for constraints to be placed on the Pop~III IMF.            
While a small fraction of DM haloes with masses as high as $\sim$ 10$^9$ M$_{\odot}$ at redshifts $z$ $\la$ 4 are not enriched due to 
star formation in their progenitors, external metal enrichment due to galactic winds is likely to preclude Pop~III star formation in a large
fraction of otherwise unenriched haloes, perhaps even preventing star formation in pristine haloes altogether after reionization is complete at $z$~$\sim$~6.   
\end{abstract}


\section{Introduction}
How long did the epoch of primordial star formation last?   
This is an important and timely question, as a primary goal of next generation telescopes, such as the 
{\it James Webb Space Telescope} (JWST; e.g. Gardner et al. 2006; Windhorst et al. 2006) and the {\it European Extremely Large Telescope}\footnote{http://www.eso.org/sci/facilities/eelt} 
(E-ELT), is to detect primordial galaxies and star clusters in the early Universe.  While these telescopes may allow to detect individual supernovae (SNe) exploding within the first 
Population (Pop)~III star clusters and galaxies (see e.g. Wise \& Abel 2005; Haiman 2008), the emission from the stellar populations within the first dwarf galaxies at $z$ $\ga$ 10 is 
likely too dim for even the JWST to detect (e.g. Ricotti et al. 2008; Johnson et al. 2009).  Instead, only very rare, massive primordial galaxies at such high redshifts, 
or dwarf galaxies and star clusters forming at lower redshift may be bright enough for detection (see e.g. Ciardi \& Ferrara 2001; Scannapieco et al. 2003).  
Therefore, whether or not Pop~III star clusters can be found by next generation telescopes
critically depends on the amount of primordial star formation that takes place at relatively low redshift.

The hierarchical nature of structure formation dictates that more massive dark matter (DM) haloes are built up from smaller ones, and that more massive haloes generally have
progenitors in which star formation may take place at earlier times.  
In order for a Pop~III star cluster to form in a DM halo, star formation in its progenitor haloes may have to be completely suppressed, lest SNe enrich the primordial gas with heavy elements
(but see e.g. Jimenez \& Haiman 2006; Pan \& Scalo 2007; Dijkstra \& Wyithe 2007).  
Exposure to a persistant photoionizing background, such as takes place in reionized regions of the Universe at high redshifts, can suppress star formation in 
DM haloes with masses substantially higher than those in which the first stars (e.g. Abel et al. 2002; Bromm et al. 2002; Yoshida et al. 2003) 
and galaxies (e.g. Greif et al. 2008; Bromm et al. 2009) form, 
completely preventing the infall of gas into haloes with circular velocities up to $\sim$ 30 km~s$^{-1}$, corresponding to virial masses of $\ga$ 3 $\times$ 10$^8$
M$_{\rm \odot}$  at $z$ $\la$ 10 (e.g. Efstathiou 1992; Thoul \& Weinberg 1996; see also Kitayama \& Ikeuchi 2000; Gnedin 2000; Ciardi \& Ferrara 2005).  
If star formation in the progenitors of these more massive haloes can be 
suppressed by the photoionizing background during reionization, then these haloes can remain unenriched until the primordial gas collapses into them to form clusters of Pop~III stars 
at relatively low redshifts, e.g. at $z$~$\la$~6.
 
\begin{figure*}
\includegraphics[width=4.5in]{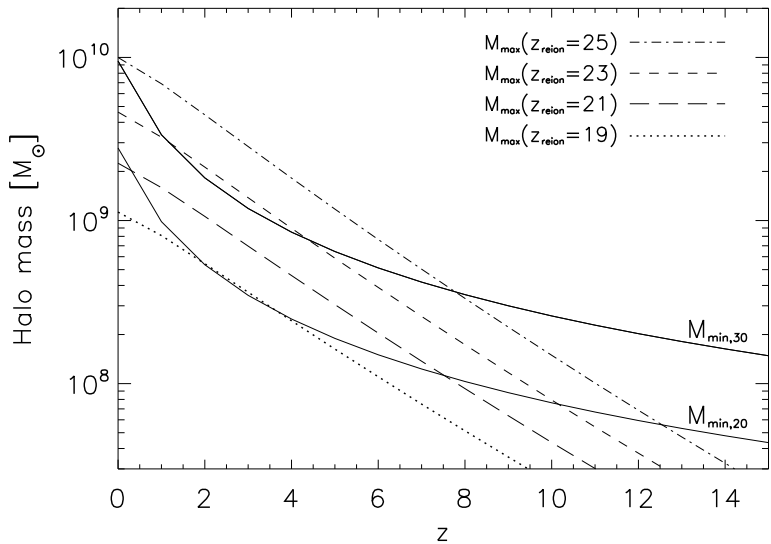}
\caption{The maximum mass $M_{\rm max}(z_{\rm reion})$ of DM haloes which have, on average, no minihalo progenitors massive enough to host star formation before a redshift $z_{\rm reion}$, 
as a function of redshift $z$.  Also shown are two possible minimum halo masses, $M_{\rm min,20}$ and $M_{\rm min,30}$, 
required for photoionized gas to collapse and form stars.  For minimum halo masses of $M_{\rm min,20}$ and $M_{\rm min,30}$,
only haloes in which star formation is suppressed by the reionization of the gas at $z_{\rm reion}$ $\ge$ 19 and $\ge$ 23, respectively, can remain pristine until the reionized gas collapses to form stars.  
These star-forming haloes may host Pop~III star clusters luminous enough to be detected in deep surveys by the JWST (see Section 4).  }
\end{figure*}

The formation of Pop~III stars at such low redshifts has been a possibility recently studied through the use of large-scale cosmological simulations. 
In their high-resolution simulations of cosmic reionization, Shin et al. (2008) (also Trac \& Cen 2007) found that haloes with masses $\sim$ 10$^8$ M$_{\odot}$ may host Pop~III star formation 
at z $\la$ 6, but their simulation did not resolve the collapse of minihaloes, i.e. haloes with virial temperature $T_{\rm vir}$ $\la$ 10$^4$ K, which could have led to the
enrichment of the gas at higher redshifts (see also Schneider et al. 2006; Tornatore et al. 2007).  In another recent study, 
Trenti et al. (2009) model the enrichment of haloes with $T_{\rm vir}$ $\ga$ 10$^4$ K by
their progenitors, and find that significant Pop~III star formation may take place at $z$ $\la$ 6, as well.  However, their study did not explicitly 
account for the suppression of star formation by photoionization, although, especially at redshifts $z$ $\ga$ 10 (Dijkstra et al. 2004), it is possible that 
star formation takes place in haloes with $T_{\rm vir}$ $\sim$ 10$^4$ K even in the presence of a photoionizing background.

In related analytical work, Wyithe \& Cen (2007) estimated the contribution of Pop~III stars to the reionization of the Universe, arguing that primordial star formation 
in neutral minihaloes may be suppressed until these haloes grow to be sufficiently massive, perhaps allowing for the formation of a significant fraction of 
Pop~III stars up to the end of reionization at $z$ $\sim$ 6.  In this work, it was assumed that star formation in minihaloes was prevented by the photodissociation of molecules at 
very high redshifts, e.g. $z$ $\ga$ 20 (Haiman et al. 1997).  However, more recent work has shown that molecular cooling can operate in minihaloes despite the build-up of the photodissociating
background radiation field (e.g. Wise \& Abel 2007; O'Shea \& Norman 2008; Johnson et al. 2008; Trenti \& Stiavelli 2009), providing motivation to relax this assumption and 
consider alternate scenarios for the formation of Pop~III stars at low redshifts.      

In the present work, we account for star formation in minihaloes at z $\ga$ 20, as well as for the suppression of star formation in DM haloes during reionization, 
with the aim of evaluating the abundance and detectability of Pop~III star clusters in reionized regions of the Universe.  
In Section 2, we outline the requirements for Pop~III star cluster
formation in the post-reionization Universe.  In Section 3, we calculate an upper limit for the abundance of such clusters and estimate the effect that 
large-scale metal-enriched galactic winds have in preventing their formation, while in Section 4 we discuss the prospects for their detection in future surveys.  
We conclude with a discussion of our results in Section 5.  For our calculations, we assume the latest cosmological parameters reported by the 
{\it Wilkinson Microwave Anisotropy Probe} (WMAP) collaboration (Komatsu et al. 2009).

\section{Conditions for Population III star formation after reionization}
The first stars are expected to have formed in DM minihaloes with virial temperature $\sim$ 10$^3$ K at redshifts $z$ $\ga$ 20 (see e.g. Bromm \& Larson 2004; O'Shea \& Norman 2007).  These 
stars were likely very massive, with a fraction of them ending their lives as powerful supernovae which enriched the intergalactic medium (IGM) with the first 
heavy chemical elements (e.g. Bromm et al. 2003; Kitayama et al. 2005; Greif et al. 2007; but see also Whalen et al. 2008a).  For a Pop~III star cluster to form in 
a given DM halo, the formation of stars which explode as SNe and enrich the gas may have to be completely prevented in its minihalo progenitors.  
We take this as a first condition for Pop~III star formation in the reionized Universe.  For simplicity, we make the conservative assumption that no Pop~III stars can form in the progenitors of 
a given halo in order for it to remain pristine, although in principle primordial stars which collapse to form black holes, instead of exploding as SNe, may not enrich the gas in their host haloes 
(see Fryer et al. 2001; Heger et al. 2003; Schneider et al. 2006).  
We discuss the related issue of the metal enrichment     
of haloes by large-scale winds from neighboring galaxies in Section 3.2.

A molecule-dissociating, so-called Lyman-Werner (LW), background radiation field can delay star formation in dark matter (DM) minihaloes, but the gas 
eventually collapses when the DM halo becomes massive enough, of the order of 10$^7$ M$_{\odot}$  for 
very high LW flux (Shang et al. 2009; see also Wise \& Abel 2007a; O'Shea \& Norman 2008; Mesinger et al. 2006, 2009).  Furthermore, as we show below, star formation in minihalo progenitors at $z$ $\ga$ 
19, on average, must be suppressed for haloes forming in the reionized Universe to avoid self-enrichment by these progenitors.  Recent modeling of the build-up of the LW 
background suggests that the LW flux may remain too low to prevent star formation in all but the rarest minihalos at $z$ $\ga$ 15 (Johnson et al. 2008; Dijkstra et al. 2008; 
Ahn et al. 2009; Trenti \& Stiavelli 2009).  Therefore, we expect that LW feedback alone is not sufficient to prevent the collapse of gas into the progenitors of DM haloes with 
masses $\ga$ 10$^8$ M$_{\odot}$, the mass range for haloes in which star formation can take after reionization is complete at $z$ $\sim$ 6.    

A more potent form of radiative feedback at high redshifts is the photoionizing background radiation that builds up during reionization.  In 
regions which are reionized early, at e.g. $z$ $\ga$ 15, the temperature of the photoheated gas is too high, at $\ga$ 10$^4$ K, for minihaloes to retain gas, 
and star formation can be suppressed. 
We note that for star formation to be suppressed by the photoionizing background during reionization, the minihalo must be subjected to the radiation before it 
reaches advanced stages of collapse, as an ionization front passing through a dense minihalo may act to induce star formation, not to prevent it
(e.g. Ahn \& Shapiro 2007; Whalen et al. 2008b; Susa et al. 2009).  Therefore, we make the simple, 
conservative assumption that the gas collapsing into minihalos with masses higher than $M_{\rm minihalo}$ = 10$^6~{\rm M_{\odot}}~((1+z)/10)^{-3/2}$,
corresponding to virial temperature $T_{\rm vir}$ $\sim$ 2 $\times$ 10$^3$ K (see e.g. Yoshida et al. 2003),
must be reionized before the virialization of the minihalo, in order for the collapse of the gas and star formation to be prevented.  
This assumption could be relaxed slightly if we account for the photoevaporation of the gas inside minihaloes that is already somewhat collapsed, 
but we do not expect this to dramatically affect our results (see e.g. Shapiro et al. 2004; Iliev et al. 2005; Wyithe \& Cen 2007).

A second condition for Pop~III star formation in unenriched DM haloes after reionization is that the DM haloes must be massive enough to allow the  
photoheated gas to cool and collapse.  Numerous studies have found that DM haloes
must have circular velocities $\ga$ 30 km~s$^{-1}$ in order for photoionized gas to collapse into them 
(e.g. Thoul \& Weinberg 1996; Quinn et al. 1996; Kepner et al. 1997; Kitayama et al. 2001).
This limit may be lower at redshifts $z$ $\ga$ 10, due to the lower average intensity of the ionizing background 
and to the higher densities
to which baryons may have already collapsed when the ionizing radiation turns on (Kitayama \& Ikeuchi 2000; Dijkstra et al. 2004; see also Chiba \& Nath 1994).  
However, there is no concensus on what this limiting mass is in cosmological haloes (Wyithe \& Cen 2007), and many previous 
calculations have been done assuming spherical symmetry (but see the cosmological simulations of Hoeft et al. 2006), 
which may underestimate the minimum halo mass, as asphericity can allow ionizing photons to penetrate deeper 
into haloes and reduce the effect of self-shielding (see Haiman 2008).  However, this is not necessarily so clear either, as dense filaments from which gas can accrete into 
cosmological haloes may also be self-shielding, in which case calculations in sperical symmetry would lead to an overestimate of the minimum mass.     
To account for some spread in the minimum mass required for gas to collapse into 
DM haloes, we consider two possibilities for this minimum mass, corresponding to minimum circular velocities of 20 and 30 km s$^{-1}$, given below:

\begin{equation}
M_{\rm min,20} = 9~\times~10^7~{\rm M_{\odot}}~\left(\frac{1+z}{10}\right)^{-3/2} \mbox{\ }
\end{equation} 
\begin{equation}
M_{\rm min,30} = 3~\times~10^8~{\rm M_{\odot}}~\left(\frac{1+z}{10}\right)^{-3/2} \mbox{\ .}
\end{equation}

Using the extended Press-Schechter formalism (Lacey \& Cole 1993), we calculate, as a function of redshift, 
the maximum mass $M_{\rm max}$($z_{\rm reion}$) of virialized DM haloes which have, on average, no minihalo progenitors with masses $\ga$ $M_{\rm minihalo}$ 
prior to redshift $z_{\rm reion}$; more precisely, we calculate the maximum mass of haloes which have $dN$/$d$ln$M_{\rm minhalo}$ $\le$ 1, where $N$ is the number of progenitor haloes 
at $z_{\rm reion}$ (see e.g. Bromm \& Clarke 2002).
This nomenclature is chosen to denote the redshift by which the region in which a halo forms must be reionized in order to prevent star formation in these minihalo progenitors.    

\begin{figure*}
\includegraphics[width=7.0in]{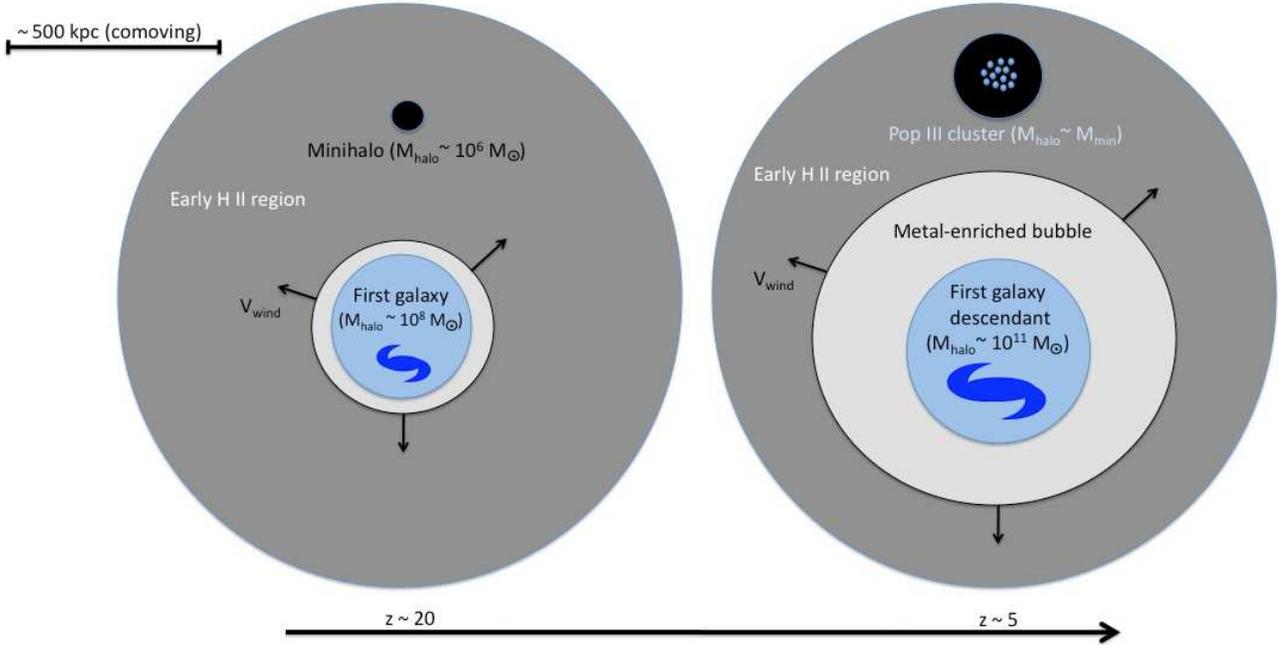}
\caption{A schematic illustration of the scenario for Pop III star cluster formation in reionized regions of the Universe described in Sections 2 and 3.  Star formation in minihaloes which are massive enough to host Pop~III star formation in the absence of radiative feedback ($\sim$ 10$^6$ M$_{\odot}$) is suppressed at redshifts $z$ $\le$ 20 by photoheating of the primordial gas within the early H~{\sc ii} region ({\it dark gray}; with a typical comoving radius of $\le$ 1 Mpc) created by a dwarf galaxy formed in a halo with a virial temperature $\ge$ 10$^4$ K ({\it light blue}; corresponding to a mass of $\sim$ 10$^8$ M$_{\odot}$ at $z$ $\sim$ 20).  The minihalo at $z$ $\sim$ 20 grows to a mass of M$_{\rm min}$, the minimum required for star formation in a photoheated gas, at $z$ $\le$ 6 (see Section 2), at which point the descendant of the first galaxy at $z$ $\sim$ 20 would typically be hosted in a $\sim$ 10$^{11}$ M$_{\odot}$ halo (see Section 4).  This galaxy is likely to drive a metal-enriched wind ({\it light gray}) into the IGM at some velocity $v_{\rm wind}$ (see Section 3.2).  If this wind does not reach the halo with mass $M_{\rm min}$ by the time stars are formed, or if the metals transported by the wind are not mixed into the halo, then the halo is assumed to host a Pop III star cluster.  Note that while the approximate size of the H~{\sc ii} region and the size of the metal-enriched bubble relative to it are roughly to scale (although shown as spherical here for simplicity), for clarity the galaxy, cluster, and halo sizes are not shown to scale.} 
\end{figure*}

We compare this maximum mass, for five different choices of $z_{\rm reion}$, to the minimum mass  
for star formation in Figure 1. Only in haloes with masses below the maximum halo mass $M_{\rm max}$($z_{\rm reion}$) and 
above the minimum mass, $M_{\rm min, 20}$ or $M_{\rm min, 30}$, at a given redshift $z$ can the reionized gas remain pristine and unenriched until it collapses to form stars.  Accordingly, 
in regions reionized at $z_{\rm reion}$ $\sim$ 25, primordial star formation may take place 
at $z$ $\la$ 12.5, on average, if the minimum halo mass is $M_{\rm min, 20}$.  If, instead, the minimum halo mass is $M_{\rm min, 30}$ then primordial star formation can 
only take place at $z$ $\la$ 8.  Also, the most massive haloes that can host metal-free star formation form at $z$ $\la$ 4 in regions that are reionized at $z_{\rm reion}$ = 19 and 23,
for a minimum halo mass of $M_{\rm min, 20}$ or $M_{\rm min, 30}$, respectively.  Indeed, due to the increase in the minimum mass for star formation with decreasing redshift, it is only in 
regions reionized at $z$ $\ga$ 19 in which Pop~III star formation can take place in reionized regions of the Universe, unless the minimum halo mass for star formation is considerably smaller 
than $M_{\rm min, 20}$.   

To summarize, two minimum requirements for the formation of a Pop~III star cluster in an unenriched DM halo after reionization, i.e. at redshifts $z$ $\la$ 6, are the following (see also Fig. 2):

\noindent(1) The halo must not have any progenitors which host star formation.  This implies a maximum halo mass $M_{\rm max}$($z_{\rm reion}$), above which star formation occurs in progenitor minihaloes
that form prior to the redshift $z_{\rm reion}$ at which the gas is reionized.  After $z_{\rm reion}$, photoheating of the gas prevents star formation, and the gas remains unenriched.

\noindent(2) The halo must be massive enough for photoheated gas to collapse into it and form stars.  This implies a minimum halo mass $M_{\rm min}$, below which stars are unable to form. 

We thus take it that only haloes with masses $\ga$ $M_{\rm min}$ and $<$ $M_{\rm max}$ can host the formation of primordial stars in reionized regions of the Universe.
A third requirement is that the haloes are not enriched by galactic winds emanating from neighboring star-forming haloes, as shown schematically in Figure 2. Such external chemical enrichment 
will limit the abundance of unenriched haloes; this is discussed further in the next Section.

\section{the abundance of Pop III Star clusters after reionization}

\subsection{An upper limit}
Now that we have the redshift and mass distribution of DM haloes which may potentially host Pop~III star formation in reionized regions of the Universe, we would like to estimate 
the abundance of such haloes as a function of redshift, thereby finding an upper limit for the abundance of Pop~III star clusters.  
As shown in Section 2, these haloes can only form in the relatively rare regions of the Universe that were reionized 
at very early times, i.e. at $z_{\rm reion}$ $\ga$ 19, meaning that these haloes are relatively rare objects.  While the early reionization history is only poorly constrained by observational 
data, there have recently been conducted a number of large-scale high-resolution simulations of this process which yield useful results for our purposes (e.g. Iliev et al. 2007; Shin et al. 2008). 
Shin et al. (2008) report their result for the volume fraction  $f_{\rm reion, vol}$($z_{\rm reion}$) of the Universe that is reionized, as a function of redshift, up to $z_{\rm reion}$ $\sim$ 22.  
For our calculation, we use the following formula, which provides a satisfactory fit to their results over the range of redshifts $z_{\rm reion}$ that we consider (i.e. 19 $\la$ $z_{\rm reion}$ $\la$ 25):  

\begin{equation}
{\rm log}(f_{\rm reion, vol}(z_{\rm reion})) = -4 + \left(\frac{21 - z_{\rm reion}}{2}\right) \mbox{\ .}   
\end{equation}

The rate of change of the collapse fraction $F_{\rm III}$($z_{\rm reion}$, $z$) of mass in haloes which may host Pop~III star formation at redshift $z$, 
if forming in regions that are reionized at $z_{\rm reion}$, is given by the rate at which mass becomes incorporated in haloes with the minimum mass for star formation, as follows:

\begin{equation}
\frac{dF_{\rm III}}{dz} = \left\{ \begin{array}{rl}
        \frac{dF(M_{\rm min},z)}{dz}  &\mbox{ if $M_{\rm min} \le M_{\rm max}(z_{\rm reion})$}, \\
  0 &\mbox{ otherwise.}
       \end{array} \right.
\end{equation}
Here $F$($M$,$z$) is the collapse fraction in DM haloes with masses $\ge$ $M$ at redshift $z$, as given by the 
Sheth-Tormen formalism (Sheth \& Tormen 1999).  $M_{\rm min}$ is the minimum mass for star formation, given either by equation (1) or (2).

As the Sheth-Tormen formalism gives the fraction of mass that is collapsed, it is 
more appropriate to use the fraction of mass $f_{\rm reion, mass}$ that is reionized than the volume-filling fraction $f_{\rm reion, vol}$ for our calculation. 
We shall assume that the mass fraction of the Universe that is 
reionized is a factor of ten higher than the volume-filling fraction $f_{\rm reion, vol}$ given in equation (3) at 
19 $\la$ $z$ $\la$ 25, due to the relatively large overdensities of the rare regions that are reionized so early. 
This is roughly consistent with the results of Shin et al. (2008), who report that at $z$ $\sim$ 20 the volume-filling fraction of H~{\sc ii} regions is of the order of 10$^{-4}$ while the 
fraction of the gas that is ionized is of the order of 10$^{-3}$, implying an average overdensity of the H~{\sc ii} regions of the order of ten (see also Furlanetto et al. 2004).  
Accounting for this fraction $f_{\rm reion, mass}(z_{\rm reion})$ of the baryonic mass in the Universe that was reionized before $z_{\rm reion}$, we integrate over $z_{\rm reion}$ 
to find the total number density of haloes virializing at redshift $z$ which are unenriched by previous star formation.  
Assuming that a single Pop~III star cluster with a lifetime of $t_{\rm cluster}$ forms in every unenriched halo, we thus estimate 
the comoving number density of Pop~III clusters as

\begin{eqnarray}
n_{\rm III}(z) & \sim & \frac{\rho_{\rm crit} \Omega_{\rm DM}}{M_{\rm min}(z)} \int_{19}^{25} \frac{dF_{\rm III}}{dz} \frac{dz}{dt} t_{\rm cluster}  \nonumber \\
             & \times &   \left|\frac{df_{\rm reion, mass}}{dz_{\rm reion}}\right| dz_{\rm reion} \nonumber \\
             & \sim &  10^{3} \; {\rm Gpc}^{-3} \left(\frac{M_{\rm min}(z)}{10^8 M_{\odot}}\right)^{-1} \left(\frac{t_{\rm cluster}}{10^6 \; {\rm yr}}\right) \left(\frac{1+z}{10}\right)^{\frac{5}{2}} \nonumber \\
             & \times &   \int_{19}^{25}    \frac{dF_{\rm III}/dz}{10^{-3}} \left|\frac{df_{\rm reion, mass}/dz_{\rm reion}}{10^{-3}}\right| dz_{\rm reion} \mbox{\ }
\end{eqnarray}
where $\rho_{\rm crit}$ is the critical density of the Universe, assuming a Hubble constant H$_{\rm 0}$ = 70.5 km~s$^{-1}$ Mpc$^{-1}$ (Komatsu et al. 2009),  
 and $\Omega_{\rm DM}$ is the fraction of this density in dark matter.  The integral is taken
from $z_{\rm reion}$ = 19 to 25, because unenriched haloes with masses $\ga$ $M_{\rm min}$ only form in regions reionized before $z_{\rm reion}$ = 19, 
as found in the previous Section, and because before $z_{\rm reion}$ = 25 
such a small fraction of the Universe is reionized that little contribution would be made to the integral for higher $z_{\rm reion}$.  
Finally, we divide by the minimum halo mass to find the number density of haloes.

\begin{figure*}
\includegraphics[width=4.12in]{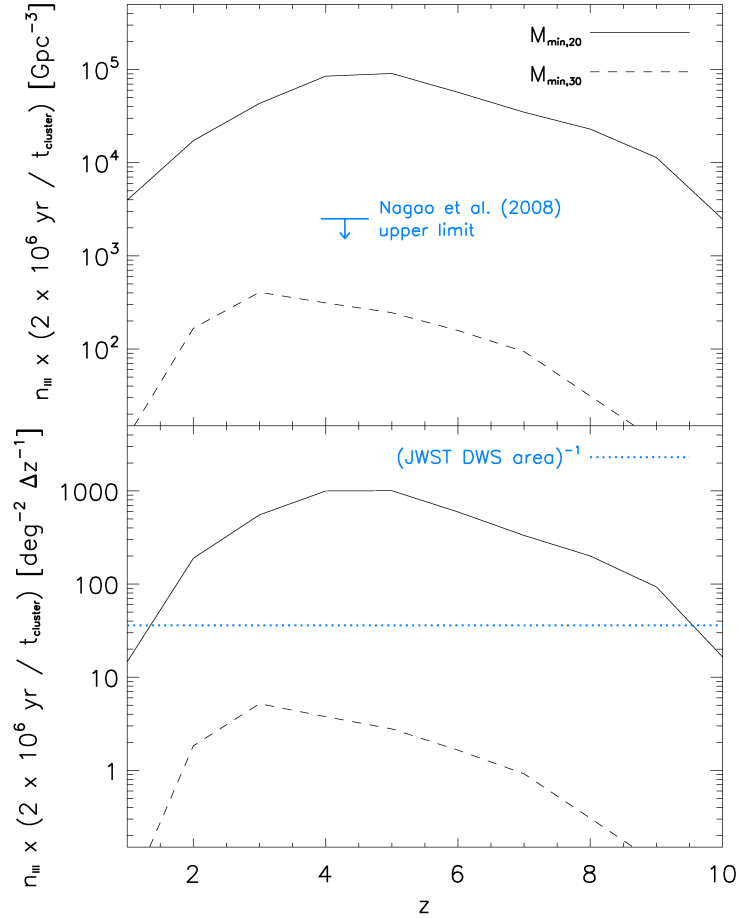}
\caption{Upper limits for the number density of Pop~III star clusters $n_{\rm III}$ formed from reionized gas,  
as given by equation (5), for our two choices 
for the minimum DM halo mass for star formation given by equations (1) and (2).  Here $n_{\rm III}$ is normalized to a cluster lifetime $t_{\rm cluster}$ = 2 $\times$ 10$^6$ yr, the expected lifetime of the most massive Pop~III stars.  {\it Top panel}: the number of clusters per comoving Gpc$^{-3}$.  Also shown is the upper limit to the 
number density of bright Pop~III star clusters at $z$ $\sim$ 4 reported by Nagao et al. (2008) (see text in Sections 3 and 4).  {\it Bottom panel}:  the number of clusters 
per redshift interval $\Delta$$z$ per square degree on the sky.    
The {\it dotted line} shows the number density of 
clusters at which there is one cluster per redshift interval, on average, within the 100 arcmin$^2$ area of the JWST Deep-Wide Survey.  This survey could reveal at most $\sim$ 100
clusters out to $z$ $\sim$ 6, but this depends sensitively on the minimum DM halo mass for star formation in a reionized gas; 
if the minimum circular velocity for such haloes is 30 km~s$^{-1}$ instead of 20 km~s$^{-1}$, then such a survey would likely turn up no Pop~III clusters.  }
\end{figure*}

We plot two curves for the number density $n_{\rm III}$ of Pop~III star clusters in the top panel of Figure 3, one for each of the choices of $M_{\rm min}$, given by equation (1) and (2).
The more massive haloes with mass $M_{\rm min, 30}$ are much rarer than those with mass $M_{\rm min, 20}$ for two reasons.  
The first is that the overall abundance of DM haloes generally decreases with halo mass.  The second reason is that 
more massive haloes have minihalo progenitors at higher redshift, meaning 
that on average they form only in the regions in which star formation is suppressed by reionization at higher $z_{\rm reion}$, and which are correspondingly more rare.  
For reasons discussed in Dijkstra et al. (2004), 
the lower minimum mass may be more accurate at higher redshift, $z$ $\la$ 10, while at redshifts $z$ $\la$ 3 the higher minimum mass is likely more accurate.  If this is so,
then the comoving number density of Pop~III star clusters may remain below $\sim$ 10$^4$~Gpc$^{-3}$ at all redshifts.  

This would be consistent with the upper limit on the abundance 
of Pop~III clusters at $z$ $\sim$ 4 found by Nagao et al. (2008), shown in the top panel of Fig. 3.  These authors report the non-detection of  
strong emitters of both Ly$\alpha$ and He~{\sc ii} $\lambda$1640 within a comoving volume of 4.03 $\times$ 10$^5$ Mpc$^3$ in the Subaru Deep Field at 3.93 $\la$ $z$ $\la$ 4.01
and 4.57 $\la$ $z$ $\la$ 4.65.  This corresponds to an upper limit on the number density of Pop~III clusters of $n_{\rm III}$ $\sim$ 2500 Gpc$^{-3}$, between the upper limits 
plotted in Fig. 3 for minimum halo masses of M$_{\rm min, 20}$ and M$_{\rm min, 30}$.  While this upper limit suggests that either the minimum halo mass for star formation
is higher than $M_{\rm min,20}$ at low redshifts or that metal enrichment by galactic winds is relatively rapid (see Section 3.2), we note that the limiting flux of the survey would have allowed
to detect only the brightest clusters (see Section 4).

In the bottom panel of Fig. 3, we plot the number density of Pop~III star clusters per degree observed on the sky per redshift interval $\Delta$$z$.  
We also plot the inverse of the 100 arcmin$^2$ area of the planned Deep-Wide Survey (DWS) to be conducted by the JWST, in which Pop~III star 
clusters may be detected (Gardner et al. 2006); this corresponds to the number density of clusters that is required for there to be at least one
cluster within this survey area, on average, per unit redshift.  The number of clusters that the JWST may detect is clearly very sensitive to the minimum halo mass for star formation; 
most optimistically, for $M_{\rm min, 20}$, there may be up to $\sim$ 100 clusters out to $z$ $\sim$ 6 per DWS area.  However, for a minimum halo mass of $M_{\rm min, 30}$ the total number density is
too low for the detection of even a single cluster in such a survey.  A larger survey area of, for example, 17 $\times$ 17 arcmin$^2$ for the E-ELT 
(ESO ELT Science Working Group\footnote{http://www.eso.org/sci/facilities/eelt/docs/ELT-SWG-apr30-1.pdf}) would, of course, provide a better chance 
to detect Pop~III clusters.   
        
As shown in equation (5), the number density of Pop~III clusters is linearly proportional to the lifetime $t_{\rm cluster}$ of an individual cluster.  Thus 
one impediment to the detection of Pop~III clusters is their brief lifetime.  Because a halo becomes enriched with metals once the first SNe explode, the lifetime of a 
purely primordial cluster is in part determined by the lifetime of the massive stars which explode as SNe, which is, at most, of the order of 10$^7$ yrs (Schaerer 2002).  
However, the timescale for the formation of second-generation, Pop~II stars from the enriched gas is also dependent on the timescale for the mixing of metal-enriched SNe ejecta
with the primordial gas, which could be $\ga$ 10$^8$ yr (see e.g. de Avillez \& Mac Low 2002).  In this case, the timescale for the formation of enriched stars would be similarly 
long; accordingly, the abundance of DM haloes hosting only Pop~III stars would be $\ga$ 50 times higher than in the fiducial case of $t_{\rm cluster}$ = 2 $\times$ 10$^6$ yr shown in Fig. 3.  
We emphasize, though, that it is the most massive stars which are the most luminous and hence it is during the lifetime of these stars, a few times 10$^6$ yrs, 
that a cluster is most easily detectable (see Section 4), meaning that the brightest clusters are also likely to be the rarest.    

While in Figure 3 we have only plotted the Pop~III cluster abundances for one value of the minimum halo mass $M_{\rm minihalo}$ = 10$^6~{\rm M_{\odot}}~((1+z)/10)^{-3/2}$ 
required for star formation in minihalos, this minimum halo mass is an uncertain quantity and may be higher if a persistant LW 
background radiation field is present (e.g. O'Shea \& Norman 2008; Trenti \& Stiavelli 2009).  If $M_{\rm minihalo}$ is indeed higher, then the abundance of Pop~III clusters
would be higher as well, since these potentially star-forming progenitor minihaloes would form at lower redshifts when $f_{\rm reion, mass}$ is considerably higher.  For a minimum minihalo mass 
an order of magnitude higher, $M_{\rm minihalo}$ = 10$^7~{\rm M_{\odot}}~((1+z)/10)^{-3/2}$, we find that the upper limits on the abundance of Pop~III star clusters may be roughly 
an order of magnitude higher than those shown in Fig. 3.     

\begin{figure*}
\includegraphics[width=4.5in]{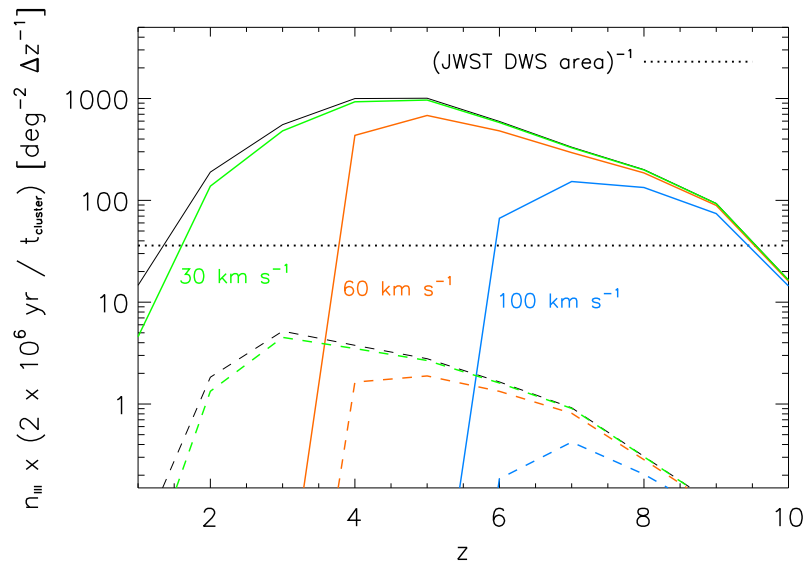}
\caption{The number density $n_{\rm III}$ of Pop~III star clusters, 
taking into account enrichment of the IGM by galactic winds.  As in Fig. 2, the solid lines correspond to a minimum halo circular velocity 
of 20 km~s$^{-1}$, while the dashed lines correspond to a minimum of 30 km~s$^{-1}$.  For each series of lines, the top ({\it black}) line is our model which neglects external metal enrichment 
(same as in Fig. 3), the {\it green} line corresponds to a wind velocity $v_{\rm wind}$ = 30 km~s$^{-1}$, the {\it red} line to $v_{\rm wind}$ = 60 km~s$^{-1}$, and the {\it blue} line
to $v_{\rm wind}$ = 100 km~s$^{-1}$.  For wind velocities of $\sim$ 60 km~s$^{-1}$, the volume-filling fraction of metals in the early H~{\sc ii} regions in which Pop~III clusters may later form reaches unity
at $z$ $\sim$ 3, thereby shutting off Pop~III star formation altogether.  For wind velocities higher than 100 km~s$^{-1}$, this occurs at $z$ $\ge$ 5.  
}
\end{figure*}

\subsection{External metal enrichment}
We have so far considered the formation of star clusters in DM haloes which are unenriched by star formation in their progenitors.  However, 
metal enrichment can also take place via galactic winds launched 
by neighboring star-forming galaxies (e.g. Nath \& Trentham 1997; Mac Low \& Ferrara 1999; Cen \& Bryan 2001; Madau et al. 2001; Mori et al. 2002; Scannapieco et al. 2003; Bertone et al. 2005; Tornatore et al. 2009).  
As such external metal enrichment 
was ignored in the last Section, our results so far for the number density of Pop~III clusters are only upper limits.  
To estimate the extent to which galactic winds enrich the IGM and thereby prevent Pop~III star formation, we compute the volume-filling fraction $f_{\rm wind}$ 
of the IGM that is reionized at $z$ $\ga$ 19 which these winds sweep up, as a function of redshift.  

As we are interested in the metal enrichment of reionized regions, we assume that winds only originate from DM haloes with masses greater than the minimum mass for star 
formation in reionized regions.  While haloes with masses $\sim$ $M_{\rm min}$ will be the much more abundant than more massive haloes (see e.g. Reed et al. 2007), 
the winds emanating from the galaxies which form in these haloes are likely to be rather weak, as the star formation efficiency in these haloes is low (e.g. Furlanetto \& Loeb 2003; Samui et al. 2008), 
due to the haloes retaining only a rather small fraction ($\sim$ 10 - 20 percent) 
of their gas due to photoheating (e.g. Thoul \& Wienberg 1996; see also Shen et al. 2009).  However, the regions of interest to us are in close proximity to the descendants of the first galaxies which reionized
their surroundings at $z_{\rm reion}$ $\ga$ 20.  As these first galaxies formed in haloes with masses of $\ga$ 5 $\times$ 10$^7$ M$_{\odot}$ at $z$ $\ga$ 20 (Trac \& Cen 2007), their descendants, on average, 
are rare haloes with masses $\ga$ 10$^{10}$ M$_{\odot}$ at $z$ $\la$ 10, as found using the extended Press-Schechter formalism.  
These galaxies likely began enriching the IGM already at $z$ $\sim$ 20, and because star formation is likely to proceed 
with a high efficiency within such massive haloes, we expect that these galaxies launch the metal-enriched winds which are principally responsible for enriching the otherwise pristine 
haloes we are considering.  Thus, under this assumption, as described schematically in Figure 2, it is the same galaxies which reionize the primordial gas at early times, thereby suppressing star formation and metal
enrichment, that are also the sources of metal-enriched winds which prevent Pop~III star formation at later times.

To model this enrichment process in a simple way, we assume that a single galaxy launches a wind from the center of each H~{\sc ii} region at redshift $z$ $\sim$ 20, and we take it 
that each H~{\sc ii} region encloses a comoving volume $V_{\rm HII}$ = 1 Mpc$^3$~$h^{-3}$ at this redshift.  While Shin et al. (2008) find that the abundance of cosmological H~{\sc ii} regions  
having comoving volumes of $\la$ 0.03 Mpc$^3$~$h^{-3}$ is $\sim$ 10 times higher than these larger H~{\sc ii} regions at $z$ $\sim$ 21, overall the larger H~{\sc ii} regions contain the majority of the reionized volume; thus, we take the larger volume as a typical value.  We assume that the metal-enriched wind propagates outward from the central galaxy at a velocity $v_{\rm wind}$, and we compute the comoving distance $R_{\rm wind}$ that it travels, as a function of redshift $z$:

\begin{equation}
R_{\rm wind}(z) = \int_{20}^z v_{\rm wind} (1+z') \frac{dt}{dz'}dz' \mbox{\ .}
\end{equation}   
Finally, we estimate the fraction $f_{\rm wind}$ of the H~{\sc ii} regions that are enriched by the winds as

\begin{equation}
f_{\rm wind}(z) \sim \frac{4\pi}{3}\frac{R_{\rm wind}^3}{ V_{\rm HII}} \mbox{\ ,}
\end{equation}
where we assume that the H~{\sc ii} regions each enclose a comoving volume of $V_{\rm HII}$ = 1 Mpc$^3$~$h^{-3}$, as explained above.
To calculate the number density of Pop~III star clusters taking into account this metal enrichment by winds, 
we take it that a fraction $f_{\rm wind}$ of haloes that are not self-enriched are instead enriched by winds; 
thus, the number density of metal-free star clusters forming in reionized regions $n_{\rm III}(z)$, as defined in equation (5), is 
reduced by this factor.  Note that for simplicity we have assumed that each individual H~{\sc ii} region has the same size and launches a wind starting at the same redshift; thus the 
metal-enriched wind volume-filling fraction of an individual H~{\sc ii} region, given by equation (7), is equal to the global fraction in this calculation. 

The resulting number densities are shown in Figure 4.  The velocity at which metal-enriched winds propagate into the IGM at high redshift is not well-constrained, 
and so we plot the number density of Pop~III clusters that we obtain for several choices, $v_{\rm wind}$ = 30, 60, and 100 km~s$^{-1}$.  For wind velocities of $\sim$ 30 km~s$^{-1}$, 
the total number of clusters within the DWS area is only slightly lower than in the case of no external metal enrichment.  For higher velocities the expected number of clusters which 
may be detected drops dramatically, with the formation of Pop~III clusters being prevented altogether below redshifts $z$ $\sim$ 3 and 5, for wind velocities of 60 and 100 km~s$^{-1}$, respectively. 
To take a fiducial case, Furlanetto \& Loeb (2003) find that the average wind velocity from the relatively massive galaxies that we are considering may be $\sim$ 60 km~s$^{-1}$, 
in which case Pop~III cluster formation is unlikely to take place at $z$ $\la$ 3, although the abundance of Pop~III clusters above this redshift is only a factor 
$\la$ 2 below the upper limits presented in Section 3.1. 

We emphasize that the simple estimates presented here do not take into account processes such as the mixing of metals with the primordial gas in the 
complex cosmological density field or the anisotropic expansion of early 
cosmological H~{\sc ii} regions. In particular, we note that the mixing of metals with the gas contained in pristine haloes may be incomplete, leaving the star-forming
gas in the centers of the haloes pristine, even if the halo as a whole is overtaken by a metal-enriched wind (see Cen \& Riquelme 2008; see also Wyithe \& Cen 2007).  
Furthermore, the metallicity of the gas in primordial haloes that is mixed with metals may not exceed the threshold necessary for the transition to low-mass, Pop~II star formation, generally taken 
to be between 10$^{-6}$ and 10$^{-3.5}$ Z$_{\odot}$ (e.g. Bromm \& Loeb 2003; Omukai et al. 2005; Frebel et al. 2007).  In this case, the stellar initial mass function 
(IMF) in these clusters may be very similar to the purely metal-free case. However, 
observations of high redshift quasars suggest that much of the enriched IGM at $z$~$\la$~5 has metallicity $Z$ $\ga$ 10$^{-4}$ Z$_{\odot}$ (Songaila 2001; Pettini et al. 2003), 
such that chemical enrichment by galactic winds may indeed effect the transition from a Pop~III to a Pop~II IMF
(see also Ferrara et al. 2000; Scannapieco et al. 2003).  In principle, though, star formation could also be 
prevented entirely in haloes with masses $\sim$ M$_{\rm min}$ due to the stripping of gas by winds with velocities exceeding the escape velocity of the halo (see Scannapieco et al. 2000).

\begin{figure*}
\includegraphics[width=7.in]{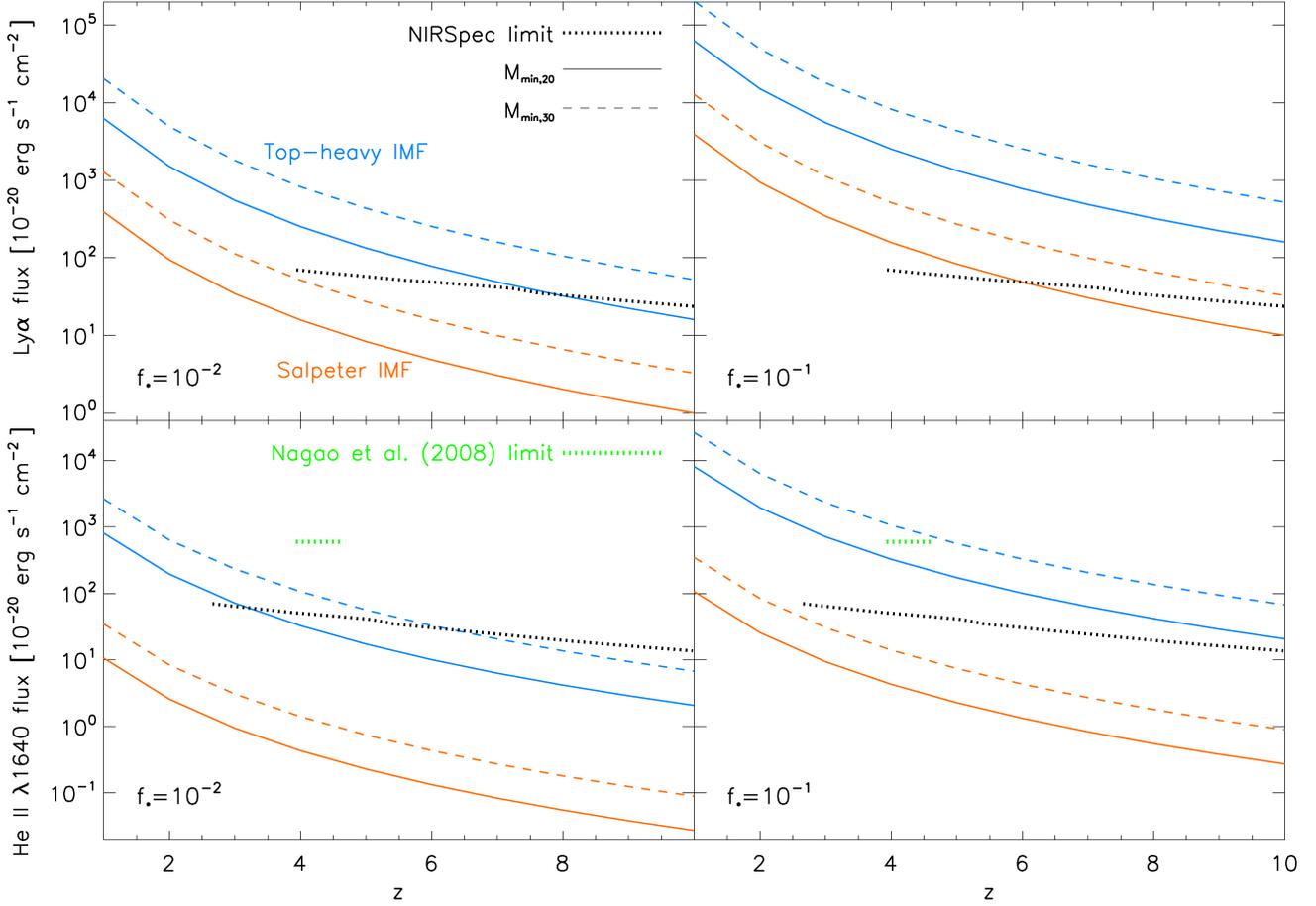}
\caption{
The flux in Ly$\alpha$ ({\it top panels}) and He~{\sc ii}~$\lambda$1640 ({\it bottom panels}) from Pop~III star clusters.  As in Figs. 2 and 3, the solid lines correspond to a minimum halo circular velocity 
of 20 km~s$^{-1}$, while the dashed lines correspond to a minimum of 30 km~s$^{-1}$; for each of the minimum halo masses, the top ({\it blue}) curves corresponds to a top-heavy IMF, while the bottom ({\it red}) 
curves corresponds to a Salpeter IMF.  The {\it left panels} show the flux of clusters assuming a star formation efficiency $f_{\rm *}$ = 10$^{-2}$, 
while the {\it right panels} correspond to the case of $f_{\rm *}$ = 10$^{-1}$.  
The other parameters characterizing the clusters are given by our fiducial choices of $f_{\rm esc}$ = 0 and $f_{\rm coll}$ = 0.1.  
The {\it black dotted} lines in each panel show the flux limits for a 3 $\sigma$ detection in a 100 hr exposure for NIRSpec ($R$=100 mode), which will be aboard the JWST. 
Ly$\alpha$ emission will appear in the wavelength range covered by NIRSpec at $z$ $\ge$ 4, 
while He~{\sc ii}~$\lambda$1640 emission will fall in this wavelength at $z$ $\ge$ 2.7.  
Detection of Ly$\alpha$ would be feasible for all clusters with a top-heavy IMF out to at least $z$ $\sim$ 8, and for clusters with a Salpeter IMF and a high star formation efficiency out to at least 
$z$ $\sim$ 6.  However,
detection of He~{\sc ii}~$\lambda$1640 is only feasible for clusters with a top-heavy IMF.  The {\it green dotted} lines in the bottom panels 
show the flux limits of the Nagao et al. (2008) survey for He~{\sc ii}~$\lambda$1640
emitters at $z$ $\sim$ 4; only the most luminous clusters would have been detectable in this survey.
}
\end{figure*}

\begin{figure*}
\includegraphics[width=7.in]{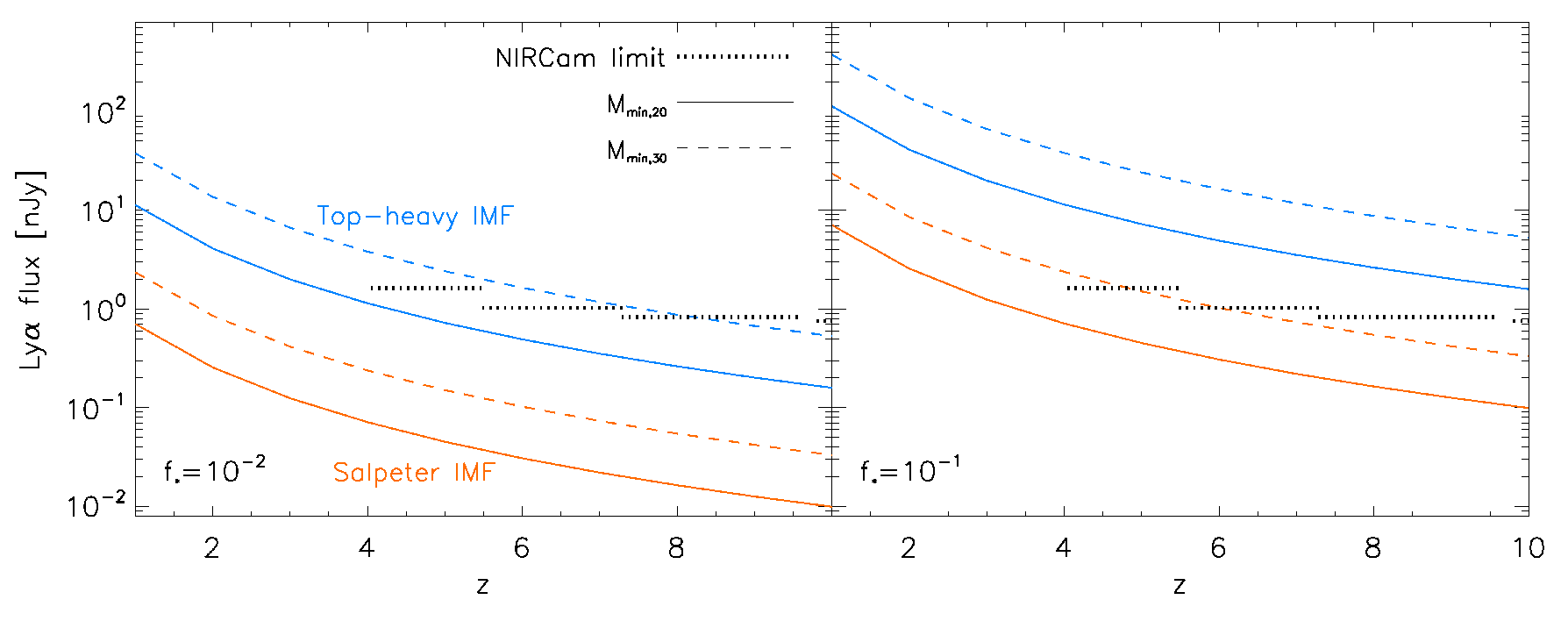}
\caption{The monochromatic Ly$\alpha$ flux of Pop~III clusters as would be observed by the NIRCam instrument in the JWST DWS, as given by equation (11), for different choices of the IMF, 
star formation efficiency $f_{\rm *}$, and minimum halo mass $M_{\rm min}$, as described in Fig. 4.  The {\it black dotted} lines are the 3 $\sigma$ detection limits expected for the DWS, 
for an exposure time of 2 $\times$ 10$^5$ seconds; the different line segments correspond to the different wide filters the wavelength range of which contains Ly$\alpha$ at redshift $z$.  
For a star formation efficiency of $f_{\rm *}$ = 10$^{-2}$ ({\it left panel}) only clusters with a top-heavy IMF may be detected, while for a high star formation efficiency $f_{\rm *}$ 
= 10$^{-1}$ ({\it right panel}) even clusters with a Salpeter IMF may be detectable out to $z$ $\sim$ 6. 
}
\end{figure*}

\section{the Detectability of pop iii star clusters after reionization}

While we have shown that there may exist a population of metal-free haloes in which Pop~III star clusters form even after reioinization, the detectability
of these clusters depends on their luminosities and spectral properties.  We consider the luminosities of the stellar clusters in 
two emission lines,  Ly$\alpha$ and He~{\sc ii}~$\lambda$1640, prominent recombination lines emitted by ionized hydrogen and
doubly ionized helium, respectively.  The ratio of the luminosities of these lines, as well as their equivalent widths, are
sensitive to the stellar IMF and metallicity.  In particular, for a given stellar mass, metal-free stars are expected to have higher surface temperatures than metal-enriched stars, 
leading to a relatively high ratio of He~{\sc ii}-ionizing photons to H~{\sc i}-ionizing photons being emitted.  In turn, this leads to relatively strong nebular 
emission in He~{\sc ii}~$\lambda$1640 (e.g. Tumlinson \& Shull 2000; Bromm et al. 2001; Oh et al. 2001; Schaerer 2002, 2003).  Detection of both Ly$\alpha$ 
and He~{\sc ii}~$\lambda$1640 can thus, in principle, facilitate the identification of Pop~III star clusters (e.g. Tumlinson et al. 2001; Nagao et al. 2008; Prescott et al. 2009).  
More massive stars also emit higher numbers of He~{\sc ii}-ionizing photons,
meaning that detection of these lines can also allow to place constraints on the stellar IMF.

While in neutral regions of the IGM Ly$\alpha$ is easily scattered (e.g. Loeb \& Rybicki 1999), 
the star clusters that we consider reside in regions which are already reionized, in principle yielding the IGM
optically thin to Ly$\alpha$, although even a small neutral fraction can result in substantial absorption (e.g. Gunn \& Peterson 1965).  In this case, other emission lines 
from ionized hydrogen will still be directly detectable, H$\alpha$ being the strongest of these lines with a luminosity $\sim$ 10 times lower than that of Ly$\alpha$ 
(e.g. Osterbrock \& Ferland 2006), and these lines could alternatively be used to constrain the metallicity and IMF (see e.g. Oh et al. 2001).  For our purposes, we will assume that the IGM is
optically thin to Ly$\alpha$ when evaluating the detectability of this emission line.

We consider Pop~III star clusters that are formed promptly, within the $\sim$ 2 Myr lifetime of the most massive individual stars (see also Oh et al. 2001), and we similarly 
evaluate the detectability of these clusters within $\sim$ 2 Myr of their formation, as it is during this time that a cluster is at its most luminous.    
To estimate the mass of the stellar clusters we take it that a fraction $f_{\rm *}$ of the gas that is able to collapse into a DM halo is converted into stars; this quantity is not well-constrained, but 
is likely of the order of a few percent (e.g. Krumholz \& Tan 2007).
In turn, we assume that only a fraction $f_{\rm coll}$ $\sim$ 0.1 of the gas associated with a DM halo is able to cool and collapse (i.e. that the collapsed baryon fraction in the halo is 10 percent of its value in the absence of photoheating), as the star-forming haloes under consideration have just grown massive 
enough to allow the photoionized gas to collapse for the first time, meaning that the majority of the gas stays photoionized and at low densities (e.g. Thoul \& Weinberg 1996; Dijkstra et al. 2004).     
We thus estimate the luminosity emitted by a Pop~III cluster in Ly$\alpha$ as 

\begin{eqnarray}
L_{\rm Ly\alpha}  & \sim  &  10^{40} \; {\rm erg} \; {\rm s}^{-1}  \left(\frac{1-f_{\rm esc, HI}}{1} \right) \left(\frac{f_{\rm *}}{0.01}\right) \left(\frac{f_{\rm coll}}{0.1}\right) \nonumber \\
               & \times & \left(\frac{Q_{\rm HI}}{10^{47} \; {\rm s}^{-1} {\rm M_{\odot}}^{-1}} \right) \left(\frac{M_{\rm min}}{10^8 {\rm M_{\odot}}}   \right)   \mbox{\ ,}
\end{eqnarray}
where $f_{\rm esc, HI}$ is the fraction of hydrogen-ionizing photons that escape the halo and so do not contribute to the nebular emission from the cluster.
The number of hydrogen-ionizing photons emitted per second per unit mass in stars, $Q_{\rm HI}$, depends on the IMF, and we consider two possible values:      
$Q_{\rm HI}$ = $10^{47} \; {\rm s}^{-1} \; {\rm M_{\odot}}^{-1}$ and ${1.6 \times 10^{48}} \; {\rm s}^{-1} \; {\rm M_{\odot}}^{-1}$, for 
a Salpeter Pop~III IMF and a top-heavy Pop~III IMF with a characteristic stellar mass $\ga$ 100 M$_{\odot}$, respectively (Tumlinson \& Shull 2000; Bromm et al. 2001).
As well, we have accounted for the fact that only a fraction of $\sim$ 2/3 of the ionizing photons that are absorbed are converted into Ly$\alpha$ photons (e.g. Spitzer 1978).  Following the same steps, we have for the luminosity in He~{\sc ii}~$\lambda$1640 

\begin{eqnarray}
L_{\rm 1640}  & \sim  &  4 \times 10^{38} \; {\rm erg} \; {\rm s}^{-1}  \left(\frac{1-f_{\rm esc, HeII}}{1} \right) \left(\frac{f_{\rm *}}{0.01}\right) \left(\frac{f_{\rm coll}}{0.1}\right) \nonumber \\
               & \times & \left(\frac{Q_{\rm HeII}}{5 \times 10^{45} \; {\rm s}^{-1} {\rm M_{\odot}}^{-1}} \right) \left(\frac{M_{\rm min}}{10^8 {\rm M_{\odot}}}   \right)   \mbox{\ ,}
\end{eqnarray}
where $f_{\rm esc, HeII}$ is the escape fraction of He~{\sc ii}-ionizing photons.
For the number of He~{\sc ii}-ionizing photons emitted per second per unit mass in stars,
we have $Q_{\rm HeII}$ = ${5 \times 10^{45}} \; {\rm s}^{-1} \;
{\rm M_{\odot}^{-1} }$ and ${3.8 \times 10^{47}} \; {\rm s}^{-1} \; {\rm M_{\odot}^{-1} }$, again for 
a Salpeter Pop~III IMF and a top-heavy Pop~III IMF, respectively (Tumlinson \& Shull 2000; Bromm et al. 2001).
The flux in Ly$\alpha$ observed at $z$ = 0 is then given by     

\begin{eqnarray}
f_{\rm Ly\alpha} &   =   &  \frac{L_{\rm Ly\alpha}}{4 \pi D_{\rm L}^2}  \nonumber \\ 
               & \sim & 10^{-20} \; {\rm erg} \; {\rm s}^{-1} \; {\rm cm}^{-2}  \left(\frac{L_{\rm Ly\alpha}}{10^{40} \; {\rm erg} \; {\rm s}^{-1}}\right) \left(\frac{1+z}{10}  \right)^{-2}  \mbox{ \ }
\end{eqnarray}
where $D_{\rm L}(z)$ is the luminosity distance to redshift $z$.  The flux in He~{\sc ii}~$\lambda$1640 is computed similarly, by replacing the Ly$\alpha$ luminosity with the
He~{\sc ii}~$\lambda$1640 luminosity.

In Figure 5 we show the fluxes in Ly$\alpha$ and He~{\sc ii}~$\lambda$1640 from clusters forming in haloes with mass $\sim$ $M_{\rm min}$, for each of our choices of this minimum halo mass, and 
for our two choices for the IMF.  We also plot the flux limits for 3 $\sigma$ detection of these emission lines with 100 hours of observation by the 
Near-Infrared Spectrograph (NIRSpec) that 
will be aboard the JWST.  NIRSpec will operate at observed wavelengths from 0.6 to 5 $\mu$m (e.g. Gardner et al. 2006). 
The lower limit of this wavelength range yields NIRSpec sensitive only to Ly$\alpha$ and He~{\sc ii}~$\lambda$1640 at redshifts $z$ $\ga$ 4 and $z$ $\ga$ 2.7, respectively. 

For the most luminous clusters that we consider, the case of a large 
fraction $f_{\rm *}$ = 10$^{-1}$ of the collapsed gas going into stars with a top-heavy IMF, both Ly$\alpha$ and 
He~{\sc ii}~$\lambda$1640 would be detectable by NIRSpec out to beyond redshift $z$ = 10, as shown in the panels on the right in Fig. 5.  
Indeed, the detection of these emission lines is much more feasible for the case of a top-heavy IMF, for all combinations 
of minimum halo mass and star formation efficiency $f_{\rm *}$.  
In particular, because even a 3 $\sigma$ detection of He~{\sc ii}~$\lambda$1640 is not feasible if clusters have a Salpeter IMF, as shown 
in the bottom panels of Fig. 5, detection of this emission line would suggest at least a somewhat top-heavy IMF.
We note, however, that the escape fractions of both H~{\sc i}- and He~{\sc ii}-ionizing photons are likely to be higher for the case of a top-heavy IMF, owing to the strong 
hydrodynamic response of the gas to photoheating (e.g. Johnson et al. 2009; see also Wise \& Cen 2009; Razoumov \& Sommer-Larsen 2009), with a corresponding decrease in the emitted 
luminosity.  Even so, the flux in He~{\sc ii}~$\lambda$1640 will likely still, in general, be higher for a more top-heavy IMF (Johnson et al. 2009).  

In the bottom panels of Fig. 5 we also plot the approximate flux limit of the survey for He~{\sc ii}~$\lambda$1640 emitters at $z$ $\sim$ 4 carried out by Nagao et al. (2008).  While this
flux limit is an order of magnitude higher than that we find for NIRSpec, the non-detection of He~{\sc ii}~$\lambda$1640 emitters in this survey already places some constraint on the 
abundance of Pop~III clusters with a top-heavy IMF and a high star formation efficiency (see Section 3).  The JWST DWS will place stronger constraints on both the abundance and luminosity of Pop~III clusters.

As the DWS will be conducted using the near-infrared imaging instrument NIRCam,  we would also like to evaluate the detectability of the clusters by comparing their expected fluxes to the flux 
limits with which NIRCam will operate for the DWS.  
Given the rest-frame spectral energy distribution (SED) of the cluster, the flux that NIRCam would detect at $z$ = 0 can be computed.  
For the expected SED of a cluster of young Pop~III stars the luminosity in Ly$\alpha$ is roughly an order of magnitude greater than the stellar and nebular continuum luminosity 
in the wavelength range covered by a given NIRCam filter (assuming a small escape fraction $f_{\rm esc}$)(e.g. Bromm et al. 2001; Schaerer 2002).  For simplicity, 
we thus calculate the flux that would be detected by NIRCam assuming that it is entirely in Ly$\alpha$.
Similar to the calculation in equation (10), we thus find the monochromatic Ly$\alpha$ flux to be 
  
\begin{eqnarray}
f_{\rm Ly\alpha} & = &  \frac{L_{\rm Ly\alpha} \lambda_{\rm Ly\alpha} (1+z)R}{4 \pi c D_{\rm L}^2}\nonumber \\ 
              & \sim & 10^{-2} \; {\rm nJy} \left(\frac{L_{\rm Ly\alpha}}{10^{40} \; {\rm erg}~{\rm s}^{-1}}\right) \left(\frac{1+z}{10}  \right)^{-1}  \left(\frac{R}{4}\right) \mbox{ \ }
\end{eqnarray} 
where $R$ is the spectral resolution; for the wide filters on the NIRCam which operate at the wavelength of Ly$\alpha$, $R$~=~4~\footnote{http://www.stsci.edu/jwst/science/sensitivity/}.
Both the monochromatic Ly$\alpha$ flux and the flux limits for the planned DWS are shown in Figure 6.  While NIRCam has an exquisite sensitivity which will allow for detection of 
fluxes as low as $\sim$ 1 nJy, Pop~III clusters may only be detected in the DWS if their IMF is top-heavy or if the star formation efficiency is relatively high, i.e. of the order of $f_{\rm *}$ $\sim$ 10$^{-1}$.   
Nonetheless, that these clusters would form at lower redshift and in more massive DM haloes suggests that their detection may be still more feasible than that of
Pop~III stars in the first dwarf galaxies at $z$ $\ga$ 10 (see e.g. Johnson et al. 2009).

As discussed in Section 3.2, the fact that these clusters would only form within the H~{\sc ii} regions surrounding the first galaxies at $z$ $\ga$ 20 (which reside in haloes with masses $\ga$ 5 $\times$ 10$^7$ M$_{\odot}$; Trac \& Cen 2007) implies that they will be found preferentially in 
the vicinity of the descendants of these first galaxies, which the extended Press-Schechter formalism dictates would typically reside in DM haloes with masses of $\sim$ 10$^{11}$ M$_{\odot}$ at 3 $\la$ $z$ $\la$ 6.  In particular, according to Shin et al. (2008) the comoving radius of the largest H~{\sc ii} regions at $z$ $\ga$ 20 is $\la$ 1 Mpc $h^{-1}$.  
Therefore, a natural prediction of our model for Pop~III star formation after reionization
is that Pop~III clusters may be found within a comoving distance of $\sim$ 1 Mpc~$h^{-1}$, corresponding to an angular separation of $\la$ 40 arcsec on the sky, of the galaxies residing in 
$\sim$ 10$^{11}$ M$_{\odot}$ haloes at 3 $\la$ $z$ $\la$ 6.

\section{Summary and Discussion}
We have shown that the suppression of star formation in DM minihaloes by photoheating during the early stages of reionization can lead to the existence of
unenriched DM haloes with masses $\ga$ 10$^8$ M$_{\odot}$ at redshifts $z$ $\la$ 10 which may host Pop~III star clusters.  While the abundance of such haloes is sensitively dependent on 
the degree to which galactic winds enrich the IGM with metals, as well as on the minimum halo mass for star formation in the reionized regions of the Universe, 
there may be up to thousands of bright Pop~III star clusters per square degree at redshifts 3 $\la$ $z$ $\la$ 6, corresponding to $\sim$ 100 
clusters within the area of the DWS to be carried out by the JWST.  This is, however, likely a strong upper limit to the abundance, as metal-enriched winds expelled by 
the same galaxies that reionize the primordial gas at early times may easily enrich it by these redshifts, depending on the speed of the winds and on 
how quickly the metals are mixed into the pristine haloes.  If this enrichment is sufficiently rapid, Pop~III star formation in pristine DM haloes may be completely 
prevented after reionization is complete at $z$ $\sim$ 6.      

Although the stellar mass, and so the luminosity, of Pop~III clusters formed in reionized regions of the Universe is limited 
by the fraction of the primordial gas that is first able to collapse into the DM haloes, we estimate that 
Ly$\alpha$ emission from these star clusters could be detected by the JWST 
if the stellar IMF is top-heavy or if the star formation efficiency $f_{\rm *}$ is high, i.e. of the order of 10 percent.  The He~{\sc ii}~$\lambda$1640 emission line, which is an indicator 
of stellar IMF and metallicity, may be detectable out to at least $z$ $\sim$ 3 for the case of a top-heavy IMF, perhaps allowing to place constraints on the Pop~III IMF, at least at low redshift.   
Furthermore, in reionized regions Pop~III clusters may form preferentially 
within $\sim$ 40 arcsec of more luminous galaxies at 3 $\la$ $z$ $\la$ 6, the descendants of the first galaxies which began reionization at $z$ $\ga$ 20, perhaps aiding
in the identification of these clusters in future surveys.        

Current observations already yield some constraints on the abundance of Pop~III star clusters at $z$ $\sim$ 4 (Dawson et al. 2004; Nagao et al. 2008; Wang et al. 2009; see also e.g. Fosbury et al. 2003; Shapley et al. 2003; Nagao et al. 2005; Jimenez \& Haiman 2006;
Prescott et al. 2009; Bouwens et al. 2009).  The non-detection of strong He~{\sc ii} $\lambda$1640 emitters at $z$ $\sim$ 4 reported by Nagao et al. (2008) yields 
an upper limit on the number density of Pop~III clusters of $n_{\rm III}$ $\sim$ 2500 Gpc$^{-3}$, between the upper limits 
plotted in Fig. 3.  However, the limiting flux in this survey is only low enough to rule out the presence of the most luminous Pop~III clusters.  
Nonetheless, this imposes a useful constraint, and suggests that 
if clusters of massive Pop~III stars do form after reionization, they may in fact form in relatively low-mass haloes in which star formation is inefficient.

We note that the Pop~III IMF in star clusters formed after reionization is likely different than that of the first stars formed in minihaloes at $z$ $\ga$ 20 (so-called Pop~III.1 stars), due to 
the onset of turbulence in the more massive haloes in which the clusters form (Wise et al. 2007b; Greif et al. 2008), to the possible enhanced abundances of the coolants H$_{\rm 2}$ and deuterium hydride (HD) in the 
reionized primordial gas (e.g. Nagakura \& Omukai 2005; Johnson \& Bromm 2006; Yoshida et al. 2007; McGreer \& Bryan 2008), and to the lower temperature of the cosmic microwave background (e.g. Larson 1998; Tumlinson 2007;
Smith et al. 2009).  
Instead of the characteristic stellar mass of $\sim$ 100 M$_{\odot}$ expected for Pop~III.1 stars formed 
in the earliest minihaloes, Pop~III.2 stars with a characteristic mass of the order of 10 M$_{\odot}$ may instead be detected in star clusters formed from reionized primordial gas.  Such clusters
would have spectral properties intermediate between the two IMFs, Salpeter and top-heavy, discussed in Section 4.

An important next step to better estimate the abundance of Pop~III star clusters will 
be to carry out cosmological simulations which capture the earliest stages of the inhomogeneous reionization of the Universe, as well as the chemical enrichment 
from both the first stars formed in minihaloes and large-scale galactic winds.  We note that the cosmological metal-enrichment simulations presented 
in Tornatore et al. (2007) already provide some indication that Pop~III clusters may form in rare regions after reionization.  
In addition, cosmological radiative transfer simulations may be necessary to estimate the masses, 
and so the luminosities, of stellar clusters that form in the DM haloes into which the reionized primordial gas is first able collapse in the course of hierarchical structure formation.      
Related to this, magnetic fields pervading the reionized gas may 
have an important dynamical effect on the formation of Pop~III star clusters at low redshifts (see Schleicher et al. 2009a); simulations tracking the growth and evolution of magnetic fields
in the IGM would thus also be of interest.  Furthermore, in the present work we have only considered the 'average' merger history of a potential Pop~III star-forming halo, 
ignoring the range of possible merger histories that such haloes may have. This assumption would clearly be relaxed within the context of a cosmological simulation.

Although we have accounted for the higher average density of regions which are reionized at early times with our choice of $f_{\rm reion, mass}$ in equation (5), we have otherwise assumed 
that the haloes in which Pop~III star clusters form are not biased, i.e. that these haloes are otherwise located randomly in space.  In principle this could lead to an overestimate of $n_{\rm III}$($z$), 
for example, if haloes with mass $M_{\rm min}$($z$) typically form in underdense regions, while it is only overdense regions which are reionized at early times.  However, haloes with mass $M_{\rm min}$($z$)
correspond to $\ga$ 1-$\sigma$ fluctuations in the cosmological density field at $z$ $\ga$ 3 (e.g. Barkana \& Loeb 2001), typically form at 3 $\la$ $z$ $\la$ 4 (see e.g. Boylan-Kolchin et al. 2009), 
and so it does not appear likely that they form preferentially in underdense regions at $z$ $\ga$ 3, the redshifts at which we find Pop~III star clusters are mostly likely to form in them.  
Therefore, we expect that accounting for the detailed clustering properties of these haloes around the descendants of the sources of reionization at $z$ $\sim$ 20 would not change 
our results substantially, although the effect of bias should be considered in more depth in future studies.      
   
Another implicit assumption throughout this work has been that stellar clusters will form in the pristine DM haloes which are massive enough for the gas to collapse into them after reionization.
There is the possibility that this gas could instead collapse to form a $\ga$ 10$^4$ M$_{\odot}$ black hole directly, if molecular cooling is suppressed (e.g. Bromm \& Loeb 2003; Begelman et al. 2006; 
Lodato \& Natarajan 2006; Spaans \& Silk 2006; Regan \& Haehnelt 2009; see also Schleicher et al. 2009b).  Shang et al. (2009) 
find that the LW background at $z$ $\la$ 8 (Ahn et al. 2009) may be high enough to suppress the formation of H$_{\rm 2}$ molecules and allow for the formation of such a massive single object.
However, it is not clear that the relatively small fraction $f_{\rm coll}$ of the gas which is able to collapse is sufficiently high for this to occur. 
Also, despite the presence of a persistant LW background, the enhanced free electron fraction in the collapsing
reionized gas will boost the formation rate of H$_{\rm 2}$ to some extent,  
perhaps aiding to prevent direct collapse to a black hole. 

Although we have explored the impact on the abundance of Pop~III clusters of different choices for the speed of metal enrichment by galactic winds, 
and for the minimum DM halo mass for star formation after reionization, we have considered only one choice for the early reionization history, based on the simulations of Shin et al. (2008).  
As the abundance of unenriched haloes is dependent on the fraction of the Universe reionized at redshifts $z$ $\ga$ 20, as well as on the typical size of cosmological H~{\sc ii} 
regions at these redshifts, our results are in principle 
quite sensitive to this choice.  Observational constraints on the early stages of reionization, for example from observations of 21-cm emission by instruments such as the 
{\it Square Kilometer Array} (e.g. Carilli et al. 2004), may thus provide indirect constraints on the abundance of unenriched haloes at lower redshifts. 
In the upcoming decade, next generation telescopes such as the JWST and the E-ELT, however, may first provide direct constraints on the amount of 
Pop~III star formation taking place in the early Universe, 
with the exciting possibility of detecting the emission from individual Pop~III star clusters, perhaps even well after the epoch of reionization.

\section*{Acknowledgements}
The author is grateful to 
Volker Bromm for the use of his extended Press-Schechter routine, and to an 
anonymous referee for comments which led to significantly 
improved clarity in the presentation of this work.  
The author also thanks Thomas Greif for valuable comments 
on an early draft, as well as 
Sadegh Khochfar and the members of the Theoretical Modeling of Cosmic Structures (TMoX) Group 
at MPE for helpful discussions.  

\end{document}